\documentclass[showpacs,preprintnumbers,amsmath,amssymb]{revtex4}

\usepackage{graphicx}
\usepackage{dcolumn}
\usepackage{bm}
\usepackage{here}

\newcommand{\bea}{\begin{eqnarray}}
\newcommand{\eea}{\end{eqnarray}}

\begin{document}
\title{Nonlinear cosmological power spectra in Einstein's gravity}
\author{Hyerim Noh}
 \email{hr@kasi.re.kr}
 \affiliation{Korea Astronomy and Space Science Institute,
              Daejon, Korea}
\author{Jai-chan Hwang}
 \email{jchan@knu.ac.kr}
 \affiliation{Department of Astronomy and Atmospheric Sciences,
              Kyungpook National University, Taegu, Korea}

\date{\today}

\begin{abstract}

Is Newton's gravity sufficient to handle the weakly nonlinear
evolution stages of the cosmic large-scale structures? Here we
resolve the issue by analytically deriving the density and velocity
power spectra to the second order in the context of Einstein's
gravity. The recently found pure general relativistic corrections
appearing in the third-order perturbation contribute to power
spectra to the second order. In this work the complete density and
velocity power spectra to the second order are derived. The power
transfers among different scales in the density power spectrum are
estimated in the context of Einstein's gravity. The relativistic
corrections in the density power spectrum are estimated to be
smaller than the Newtonian one to the second order, but these could
be larger than higher-order nonlinear Newtonian terms.

\end{abstract}
\noindent 

\maketitle

%
%
\section{Introduction}

The weakly nonlinear process of gravitating system has fundamental
importance in cosmology
\cite{textbook,Zeldovich-1965,Vishniac-1983,quasilinear}. The
large-scale structure in the observable universe apparently show its
nonlinear nature. Considering the success of the Friedmann cosmology
with its spatial homogeneity and isotropy assumptions, it is likely
that in near horizon scale the structures are in near linear stage.
In the small scale, however, the cosmic structure (say, distribution
and motion of galaxies) are in fully nonlinear stage. In between the
two scales, we have weakly nonlinear stage. Even for the nonlinear
structures in the present epoch, as the cosmological structure grows
under gravity from linear to nonlinear stages there must be
transition era which can be regarded as weakly nonlinear. The
density and velocity power spectra provide the main observations of
the large-scale structure which can be directly compared with
theories of the structure formation. All theoretical studies of the
weakly nonlinear evolution of the large-scale structure have been
based on Newton's gravity
\cite{textbook,Zeldovich-1965,Vishniac-1983,quasilinear}. Is
Newton's gravity sufficient to handle the situation? In this work we
will resolve the issue by analytically deriving the power spectra to
the second order in the context of Einstein's gravity.

We will derive the second-order density and velocity power spectra
in the context of relativistic cosmology for the first time. In a
zero-pressure medium the pure general relativistic corrections
appear in the third order; this was only recently shown by us in
\cite{third-order}. The third-order relativistic corrections,
however, contribute to the density and velocity power spectra even
to the second order which was unknown previously. We show that
compared with the Newtonian contributions to the second-order power
spectra the relativistic contributions are generically multiplied by
a factor (scale/horizon scale) squared, thus suppressed in the small
scale. In the Newtonian theory, in the second-order density power
spectrum, the nonlinear power transfer from large-scale nonlinearity
is known to exactly cancel to the leading order, thus opening a
possibility of relativistc effect becoming important. In this paper
we will show that, even in such a situation, the pure general
relativistic contribution to the density power spectrum is smaller
than the remaining second-order Newtonian ones. We conclude that,
even in the context of Einstein's gravity, up to the second order,
the $k^4$ long wavelength tail in the density power spectrum
previously known in the Newtonian analysis is the only important
effect of the nonlinear power transfer among different scales.
However, the leading order relativistic contribution in the
second-order power spectra could be larger than the higher (third
and higher) order pure Newtonian contributions to the power spectra
\cite{quasilinear}, thus demanding a caution in the pure Newtonian
study.

%
%
\section{Basic equations}

We consider an irrotational dust (zero-pressure fluid) without the
gravitational waves in a {\it flat} Friedmann background. To the
third order the basic perturbation equations in Einstein's gravity
are recently derived in \cite{third-order}. These are \bea
   & & \dot \delta + {1 \over a} \nabla \cdot {\bf u}
       = - {1 \over a} \nabla \cdot \left( \delta {\bf u} \right)
       + {1 \over a} \left[ 2 \varphi {\bf u}
       - \nabla \left( \Delta^{-1} X \right) \right] \cdot \nabla \delta,
   \label{delta-eq-3rd} \\
   & & {1 \over a} \nabla \cdot \left( \dot {\bf u}
       + {\dot a \over a} {\bf u} \right)
       + 4 \pi G \varrho \delta
       = - {1 \over a^2} \nabla \cdot \left( {\bf u}
       \cdot \nabla {\bf u} \right)
   \nonumber \\
   & & \qquad
       - {2 \over 3 a^2} \varphi
       {\bf u} \cdot \nabla \left( \nabla \cdot {\bf u} \right)
       + {4 \over a^2} \nabla \cdot \left[ \varphi
       \left( {\bf u} \cdot \nabla {\bf u}
       - {1 \over 3} {\bf u} \nabla \cdot {\bf u} \right) \right]
       - {\Delta \over a^2}
       \left[ {\bf u} \cdot \nabla \left( \Delta^{-1} X \right) \right]
       + {1 \over a^2} {\bf u} \cdot \nabla X
       + {2 \over 3a^2} X \nabla \cdot {\bf u},
   \label{u-eq-3rd} \\
   & &
       X \equiv
       2 \varphi \nabla \cdot {\bf u}
       - {\bf u} \cdot \nabla \varphi
       + {3 \over 2} \Delta^{-1} \nabla \cdot
       \left[ {\bf u} \cdot \nabla \left( \nabla \varphi \right)
       + {\bf u} \Delta \varphi \right],
   \label{X-eq-3rd}
\eea where $\delta$ and ${\bf u}$ are the density perturbation
($\delta \equiv \delta \varrho / \varrho$) and the perturbed
velocity, respectively; $a$ is the cosmic scale factor and $\varrho$
is the background density. A perturbed order variable $\varphi$ is a
metric perturbation variable related to the perturbed three-space
curvature. To the linear order $\varphi$ can be related to $\delta$
and ${\bf u}$ as \cite{third-order} \bea
   & & \varphi = {1 \over c^2} \left( - \delta \Phi
       + \dot a \Delta^{-1} \nabla \cdot {\bf u} \right),
   \label{varphi}
\eea where $\delta \Phi$ is the perturbed order Newtonian
gravitational potential; $\delta \Phi$ is related to $\delta$ by
Poisson's equation \bea
   & & \Delta \Phi = 4 \pi G \varrho a^2 \delta.
   \label{Poisson-eq}
\eea Up to the second order, remarkably, Eqs.\ (\ref{delta-eq-3rd})
and (\ref{u-eq-3rd}) {\it coincide} exactly with the ones in
Newtonian theory; we call it a relativistic/Newtonian correspondence
to the second order \cite{second-order}. This is why we simply call
$\delta$ and ${\bf u}$ as the (Newtonian) density and velocity
perturbations even in the present relativistic situation; in the
relativistic context $\delta$, ${\bf u}$ and $\varphi$ are related
to the certain gauge-invariant combinations of variables, see
\cite{second-order,third-order}. It is also remarkable to notice
that all the pure relativistic third-order correction terms have
$\varphi$ factor compared with the relativistic/Newtonian
second-order terms. Contribution of the gravitational waves to the
third order can be found in \cite{third-order}; for equations in the
multi-component case, see the third reference in \cite{third-order}.
We note that as the above equations are derived in the relativistic
perturbation theory, these are valid in the fully general
relativistic situation and in all scales (including the
super-horizon scale) as long as the perturbative assumption is met.

%
%
\section{Solutions in the phase space}

As we are considering a flat background, we may take the Fourier
transformation defined as $F ({\bf k}) = \int d^3x F ({\bf
x})e^{i{\bf k\cdot x}}$. By introducing ${\bf u} \equiv \nabla u$
and $\theta \equiv \nabla \cdot {\bf u} = \Delta u$, we have ${\bf
u} ({\bf k}, t) = - i {\bf k} u ({\bf k}, t)$ and $\theta ({\bf k},
t) = - k^2 u ({\bf k}, t)$. In the phase space, Eqs.\
(\ref{delta-eq-3rd})-(\ref{X-eq-3rd}) become \bea
   & & \dot \delta ({\bf k}, t)
       + {1 \over a} \theta ({\bf k}, t)
       = - {1 \over a} {1 \over (2 \pi)^3} \int d^3 k^\prime
       {1 \over 2} \left[ \delta ({\bf k}^\prime, t)
       \theta ({\bf k} - {\bf k}^\prime, t)
       { {\bf k} \cdot ( {\bf k} - {\bf k}^\prime ) \over
       | {\bf k} - {\bf k}^\prime |^2 }
       + C^+ \right]
   \nonumber \\
   & & \qquad
       + {2 \over a} {1 \over (2 \pi)^6} \int d^3 k^\prime
       \int d^3 k^{\prime\prime}
       {1 \over 3} \left[ \varphi ({\bf k}^\prime, t)
       \delta ({\bf k} - {\bf k}^\prime - {\bf k}^{\prime\prime}, t)
       \theta ({\bf k}^{\prime\prime}, t)
       {({\bf k} - {\bf k}^\prime - {\bf k}^{\prime\prime})
       \cdot {\bf k}^{\prime\prime} \over k^{\prime\prime 2}}
       + C^{++} \right]
   \nonumber \\
   & & \qquad
       - {1 \over a} {1 \over (2 \pi)^3}
       \int d^3 k^\prime
       {1 \over 2} \left[ { ({\bf k} - {\bf k}^\prime)
       \cdot {\bf k}^\prime \over k^{\prime 2}}
       X ({\bf k}^\prime, t)
       \delta ({\bf k}- {\bf k}^\prime, t)
       + C^+ \right],
   \label{delta-k-eq} \\
   & & {1 \over a} \left[
       \dot \theta ({\bf k}, t)
       + {\dot a \over a} \theta ({\bf k}, t) \right]
       + 4 \pi G \varrho \delta ({\bf k}, t)
       = - {1 \over a^2}
       {1 \over (2 \pi)^3} \int d^3 k^\prime
       \theta ({\bf k}^\prime, t)
       \theta ({\bf k} - {\bf k}^\prime, t)
       {1 \over 2} {k^2 \over k^{\prime 2}}
       { {\bf k}^\prime \cdot ( {\bf k} - {\bf k}^\prime )
       \over | {\bf k} - {\bf k}^\prime |^2 }
   \nonumber \\
   & & \qquad
       - {1 \over a^2} {1 \over (2 \pi)^6}
       \int d^3 k^\prime \int d^3 k^{\prime\prime}
       {1 \over 3}
       \Bigg[
       \varphi ({\bf k}^\prime, t)
       \theta ({\bf k}^{\prime\prime}, t)
       \theta ({\bf k} - {\bf k}^\prime - {\bf k}^{\prime\prime}, t)
   \nonumber \\
   & & \qquad
       \times
       \left( {2 \over 3}
       { ( 3 {\bf k} - {\bf k}^\prime - {\bf k}^{\prime\prime} )
       \cdot {\bf k}^{\prime\prime}
       \over k^{\prime\prime 2} }
       - 4 { ({\bf k} - {\bf k}^\prime - {\bf k}^{\prime\prime})
       \cdot {\bf k}^{\prime\prime}
       \over k^{\prime\prime 2} }
       { {\bf k} \cdot
       ({\bf k} - {\bf k}^\prime - {\bf k}^{\prime\prime})
       \over | {\bf k} - {\bf k}^\prime - {\bf k}^{\prime\prime} |^2
       } \right)
       + C^{++}
       \bigg]
   \nonumber \\
   & & \qquad
       + {1 \over a^2} {1 \over (2 \pi)^3}
       \int d^3 k^\prime {1 \over 2}
       \left[ { ({\bf k} - {\bf k}^\prime) \cdot {\bf k}^\prime
       \over k^{\prime 2} }
       \left( 1- {k^2 \over |{\bf k} - {\bf k}^\prime|^2} \right)
       \theta ({\bf k}^\prime, t)
       X ({\bf k} - {\bf k}^\prime, t)
       + {2 \over 3} \theta ({\bf k} - {\bf k}^\prime, t)
       X ({\bf k}^\prime, t)
       + C^+ \right],
   \label{theta-k-eq} \\
   & & X ({\bf k}, t)
       = {1 \over (2 \pi)^3} \int d^3 k^\prime
       \bigg\{
       \theta ({\bf k}^\prime, t)
       \varphi ({\bf k} - {\bf k}^\prime, t)
       \left[
       1 - {1 \over 2} { ({\bf k} - {\bf k}^\prime)
       \cdot {\bf k}^\prime \over k^{\prime 2} }
       \left(
       1 - {3 \over 2} { ({\bf k} - {\bf k}^\prime)
       \cdot {\bf k} \over k^2 }
       \right)
       + {3 \over 4} { {\bf k}^\prime \cdot {\bf k}
       \over k^2}
       { | {\bf k} - {\bf k}^\prime |^2
       \over k^{\prime 2} } \right]
       + C^+
       \bigg\}.
   \nonumber \\
   \label{X-k-eq}
\eea where $C^+$ indicates terms replacing ${\bf k}^\prime$ to ${\bf
k} - {\bf k}^\prime$; $C^{++}$ indicates two sets of terms, one
replacing ${\bf k}^\prime$ to ${\bf k} - {\bf k}^\prime - {\bf
k}^{\prime\prime}$, and the other replacing ${\bf k}^{\prime\prime}$
to ${\bf k} - {\bf k}^\prime - {\bf k}^{\prime\prime}$.

%
%

In order to derive the perturbative solutions we expand \bea
   & & \delta ({\bf k}, t) = \delta_1 ({\bf k}, t) + \delta_2 ({\bf k}, t)
       + \delta_3 ({\bf k}, t) + \dots ,
\eea and similarly expand $\theta$ and $\varphi$. To the linear
order, Eqs.\ (\ref{delta-k-eq}) and (\ref{theta-k-eq}) give $\ddot
\delta_1 + 2 (\dot a / a) \dot \delta_1 - 4 \pi G \varrho \delta_1 =
0$. Equations up to this point are valid in the presence of the
cosmological constant. In the following, in order to derive analytic
solutions we {\it assume} an absence of the cosmological constant.
In a flat background without the cosmological constant, we have $a
\propto t^{2/3}$, $6 \pi G \varrho = t^{-2}$, thus we have two
solutions $\delta_1 ({\bf k}, t) \propto t^{2/3}$ and $t^{-1}$. We
{\it ignore} the decaying solution in an expanding phase and set
\bea
   & & \delta_1 ({\bf k}, t)
       \equiv A ({\bf k}) e^{i \phi({\bf k})} t^{2/3}, \quad
       \theta_1 ({\bf k}, t)
       = - {2 \over 3} A ({\bf k}) e^{i \phi({\bf k})}
       a t^{-1/3}, \quad
       \varphi_1 ({\bf k}, t)
       = {5 \over 2}
       \left( {\ell / \ell_{H}} \right)^2
       t^{2/3}
       A ({\bf k}) e^{i \phi({\bf k})}.
   \label{linear-sol-1}
\eea In a flat background {\it without} the cosmological constant,
to the linear order, from Eqs.\ (\ref{varphi}) and
(\ref{Poisson-eq}) we have $\varphi = - (5/3) \delta \Phi / c^2$. We
introduced a scale $\ell \equiv a/k$ and the Hubble horizon scale
$\ell_H \equiv c/(\dot a/a)$; thus, ${\ell / \ell_{H}} \equiv {\dot
a / (kc)}$. Notice that $\varphi_1$ is time independent. To the
second order from Eqs.\ (\ref{delta-k-eq}) and (\ref{theta-k-eq}) we
have the solutions \cite{Vishniac-1983} \bea
   & & \delta_2 ({\bf k}, t)
       = {1 \over 14} t^{4/3}
       {1 \over (2 \pi)^3}
       \int d^3 k^\prime
       A ({\bf k}^\prime) A ({\bf k} - {\bf k}^\prime)
       e^{ i \phi({\bf k}^\prime)
       + i \phi ({\bf k} - {\bf k}^\prime) }
       J ({\bf k}, {\bf k}^\prime, {\bf k} - {\bf k}^\prime),
   \label{second-order-sol-1} \\
   & & \theta_2 ({\bf k}, t)
       = - {1 \over 21} a t^{1/3}
       {1 \over (2 \pi)^3}
       \int d^3 k^\prime
       A ({\bf k}^\prime) A ({\bf k} - {\bf k}^\prime)
       e^{ i \phi({\bf k}^\prime)
       + i \phi ({\bf k} - {\bf k}^\prime) }
       L ({\bf k}, {\bf k}^\prime, {\bf k} - {\bf k}^\prime),
   \label{second-order-sol-2}
\eea where \bea
   & &
       J ({\bf k}, {\bf k}^\prime, {\bf k} - {\bf k}^\prime)
       \equiv
       \left[
       4 F ({\bf k}, {\bf k}^\prime, {\bf k} - {\bf k}^\prime)
       + 5 H ({\bf k}, {\bf k}^\prime)
       + 5 H ({\bf k}, {\bf k} - {\bf k}^\prime)
       \right],
   \nonumber \\
   & &
       L ({\bf k}, {\bf k}^\prime, {\bf k} - {\bf k}^\prime)
       \equiv
       \left[
       8 F ({\bf k}, {\bf k}^\prime, {\bf k} - {\bf k}^\prime)
       + 3 H ({\bf k}, {\bf k}^\prime)
       + 3 H ({\bf k}, {\bf k} - {\bf k}^\prime)
       \right],
   \nonumber \\
   & &
       H ({\bf k}, {\bf k}^\prime)
       \equiv { {\bf k} \cdot {\bf k}^\prime
       \over k^{\prime 2} }, \quad
       F ({\bf k}, {\bf k}^\prime, {\bf k} - {\bf k}^\prime)
       \equiv {1 \over 2} {k^2 \over k^{\prime 2}}
       { {\bf k}^\prime \cdot ( {\bf k} - {\bf k}^\prime )
       \over | {\bf k} - {\bf k}^\prime |^2 }.
\eea In order to derive third-order solutions we need $X$ to the
second order. Using Eq.\ (\ref{linear-sol-1}), Eq.\ (\ref{X-k-eq})
becomes \bea
   & & X_2 ({\bf k}, t)
       = - {20 \over 27} {1 \over c^2 k^2}
       {1 \over (2 \pi)^3} \int d^3 k^\prime
       A ({\bf k}^\prime) A ({\bf k} - {\bf k}^\prime)
       e^{i\phi ({\bf k}^\prime) + i \phi ({\bf k} - {\bf k}^\prime)}
       M ({\bf k}, {\bf k}^\prime, {\bf k} - {\bf k}^\prime)
       a^3 t^{-5/3},
   \label{X-k-eq-4} \\
   & & M ({\bf k}, {\bf k}^\prime, {\bf k} - {\bf k}^\prime)
       \equiv
       {k^2 \over k^{\prime 2}}
       + { k^2 \over | {\bf k} - {\bf k}^\prime |^2 }
       + {3 \over 4} { {\bf k} \cdot {\bf k}^\prime
       \over k^{\prime 2} }
       + {3 \over 4}
       { {\bf k} \cdot ({\bf k} - {\bf k}^\prime )
       \over | {\bf k} - {\bf k}^\prime |^2 }
       - {1 \over 4} {k^2 \over k^{\prime 2}}
       { {\bf k}^\prime \cdot ({\bf k} - {\bf k}^\prime)
       \over | {\bf k} - {\bf k}^\prime |^2 }.
   \label{M-def}
\eea

To the third order from Eqs.\ (\ref{delta-k-eq}) and
(\ref{theta-k-eq}), using Eqs.\ (\ref{linear-sol-1})-(\ref{M-def}),
we have the solutions \bea
   & & \delta_3 ({\bf k}, t)
       =
       {1 \over 4} t^2 {1 \over (2 \pi)^6} \int d^3 k^\prime
       \int d^3 k^{\prime\prime}
       A ({\bf k}^{\prime\prime})
       e^{i \phi ({\bf k}^{\prime\prime})}
       \Bigg\{
       A ({\bf k}^\prime)
       A ({\bf k} - {\bf k}^\prime - {\bf k}^{\prime\prime})
       e^{ i \phi ({\bf k}^\prime)
       + i \phi({\bf k} - {\bf k}^\prime - {\bf k}^{\prime\prime}) }
   \nonumber \\
   & & \qquad
       \times
       \bigg[
       {2 \over 63}
       F ( {\bf k}, {\bf k}^\prime, {\bf k} - {\bf k}^\prime )
       L ( {\bf k} - {\bf k}^\prime, {\bf k}^{\prime\prime},
       {\bf k} - {\bf k}^\prime - {\bf k}^{\prime\prime} )
   \nonumber \\
   & & \qquad
       + {1 \over 18}
       H ( {\bf k}, {\bf k}^\prime )
       J ( {\bf k} - {\bf k}^\prime, {\bf k}^{\prime\prime},
       {\bf k} - {\bf k}^\prime - {\bf k}^{\prime\prime} )
       + {1 \over 18}
       H ( {\bf k}, {\bf k} - {\bf k}^\prime )
       L ( {\bf k} - {\bf k}^\prime, {\bf k}^{\prime\prime},
       {\bf k} - {\bf k}^\prime - {\bf k}^{\prime\prime} )
       \bigg]
       + C^+
       \Bigg\}
   \nonumber \\
   & & \qquad
       + {5 \over 21}
       \left( {\ell \over \ell_H} \right)^2
       t^2
       {1 \over (2 \pi)^6} \int d^3 k^\prime
       \int d^3 k^{\prime\prime}
       A ({\bf k}^{\prime\prime})
       e^{i \phi ({\bf k}^{\prime\prime})}
       \Bigg\{
       A ({\bf k}^\prime)
       A ({\bf k} - {\bf k}^\prime - {\bf k}^{\prime\prime})
       e^{ i \phi ({\bf k}^\prime)
       + i \phi({\bf k} - {\bf k}^\prime - {\bf k}^{\prime\prime})}
   \nonumber \\
   & & \qquad
       \times
       \bigg[
       {k^2 \over k^{\prime 2}}
       \left(
       - 3 { ({\bf k} - {\bf k}^\prime - {\bf k}^{\prime\prime})
       \cdot {\bf k}^{\prime\prime} \over k^{\prime\prime 2} }
       + {4 \over 3} {( {\bf k}^\prime
       + {\bf k}^{\prime\prime}) \cdot {\bf k}^{\prime\prime}
       \over k^{\prime\prime2} }
       - 4 { ({\bf k} - {\bf k}^\prime - {\bf k}^{\prime\prime})
       \cdot {\bf k}^{\prime\prime} \over k^{\prime\prime 2} }
       { {\bf k} \cdot ({\bf k} - {\bf k}^\prime - {\bf k}^{\prime\prime})
       \over | {\bf k} - {\bf k}^\prime - {\bf k}^{\prime\prime} |^2}
       \right)
       + C^{++}
       \bigg]
   \nonumber \\
   & & \qquad
       + \bigg[
       A ({\bf k}^\prime)
       A ({\bf k} - {\bf k}^\prime - {\bf k}^{\prime\prime})
       e^{ i \phi ({\bf k}^\prime)
       + i \phi({\bf k} - {\bf k}^\prime - {\bf k}^{\prime\prime})}
       M ( {\bf k} - {\bf k}^\prime, {\bf k}^{\prime\prime},
       {\bf k} - {\bf k}^\prime - {\bf k}^{\prime\prime} )
       {k^2 \over |{\bf k} - {\bf k}^\prime|^2}
   \nonumber \\
   & & \qquad
       \times
       \left(
       {1 \over 2}
       - {3 \over 2} { {\bf k} \cdot {\bf k}^\prime
       \over k^{\prime 2} }
       + {15 \over 4}
       { {\bf k}^\prime \cdot ({\bf k} - {\bf k}^\prime)
       \over |{\bf k} - {\bf k}^\prime|^2 }
       + {3 \over 2} {k^2 \over k^{\prime 2}}
       { {\bf k}^\prime \cdot ({\bf k} - {\bf k}^\prime)
       \over |{\bf k} - {\bf k}^\prime|^2 }
       \right)
       + C^+
       \bigg]
       \Bigg\},
   \label{third-order-sol-1} \\
   & & \theta_3 ({\bf k}, t)
       =
       - {1 \over 63} a t {1 \over (2 \pi)^6} \int d^3 k^\prime
       \int d^3 k^{\prime\prime}
       A ({\bf k}^{\prime\prime})
       e^{i \phi ({\bf k}^{\prime\prime})}
       \Bigg\{
       A ({\bf k}^\prime)
       A ({\bf k} - {\bf k}^\prime - {\bf k}^{\prime\prime})
       e^{ i \phi ({\bf k}^\prime)
       + i \phi({\bf k} - {\bf k}^\prime - {\bf k}^{\prime\prime}) }
   \nonumber \\
   & & \qquad
       \times
       \bigg[
       F ( {\bf k}, {\bf k}^\prime, {\bf k} - {\bf k}^\prime )
       L ( {\bf k} - {\bf k}^\prime, {\bf k}^{\prime\prime},
       {\bf k} - {\bf k}^\prime - {\bf k}^{\prime\prime} )
   \nonumber \\
   & & \qquad
       + {1 \over 4}
       H ( {\bf k}, {\bf k}^\prime )
       J ( {\bf k} - {\bf k}^\prime, {\bf k}^{\prime\prime},
       {\bf k} - {\bf k}^\prime - {\bf k}^{\prime\prime} )
       + {1 \over 4}
       H ( {\bf k}, {\bf k} - {\bf k}^\prime )
       L ( {\bf k} - {\bf k}^\prime, {\bf k}^{\prime\prime},
       {\bf k} - {\bf k}^\prime - {\bf k}^{\prime\prime} )
       \bigg]
       + C^+
       \Bigg\}
   \nonumber \\
   & & \qquad
       + {5 \over 21}
       \left( {\ell \over \ell_H} \right)^2
       a t
       {1 \over (2 \pi)^6} \int d^3 k^\prime
       \int d^3 k^{\prime\prime}
       A ({\bf k}^{\prime\prime})
       e^{i \phi ({\bf k}^{\prime\prime})}
       \Bigg\{
       A ({\bf k}^\prime)
       A ({\bf k} - {\bf k}^\prime - {\bf k}^{\prime\prime})
       e^{ i \phi ({\bf k}^\prime)
       + i \phi({\bf k} - {\bf k}^\prime - {\bf k}^{\prime\prime})}
   \nonumber \\
   & & \qquad
       \times
       \bigg[
       {k^2 \over k^{\prime 2}}
       \left(
       - {2 \over 3} { ({\bf k} - {\bf k}^\prime - {\bf k}^{\prime\prime})
       \cdot {\bf k}^{\prime\prime} \over k^{\prime\prime 2} }
       - {16 \over 9} {( {\bf k}^\prime
       + {\bf k}^{\prime\prime}) \cdot {\bf k}^{\prime\prime}
       \over k^{\prime\prime2} }
       + {16 \over 3} { ({\bf k} - {\bf k}^\prime - {\bf k}^{\prime\prime})
       \cdot {\bf k}^{\prime\prime} \over k^{\prime\prime 2} }
       { {\bf k} \cdot ({\bf k} - {\bf k}^\prime - {\bf k}^{\prime\prime})
       \over | {\bf k} - {\bf k}^\prime - {\bf k}^{\prime\prime} |^2}
       \right)
       + C^{++}
       \bigg]
   \nonumber \\
   & & \qquad
       + \bigg[
       A ({\bf k}^\prime)
       A ({\bf k} - {\bf k}^\prime - {\bf k}^{\prime\prime})
       e^{ i \phi ({\bf k}^\prime)
       + i \phi({\bf k} - {\bf k}^\prime - {\bf k}^{\prime\prime})}
       M ( {\bf k} - {\bf k}^\prime, {\bf k}^{\prime\prime},
       {\bf k} - {\bf k}^\prime - {\bf k}^{\prime\prime} )
       {k^2 \over |{\bf k} - {\bf k}^\prime|^2}
   \nonumber \\
   & & \qquad
       \times
       \left(
       - {2 \over 3}
       + 2 { {\bf k} \cdot {\bf k}^\prime
       \over k^{\prime 2} }
       - {3 \over 2}
       { {\bf k}^\prime \cdot ({\bf k} - {\bf k}^\prime)
       \over |{\bf k} - {\bf k}^\prime|^2 }
       - 2 {k^2 \over k^{\prime 2}}
       { {\bf k}^\prime \cdot ({\bf k} - {\bf k}^\prime)
       \over |{\bf k} - {\bf k}^\prime|^2 }
       \right)
       + C^+
       \bigg]
       \Bigg\}.
   \label{third-order-sol-2}
\eea We note that the pure general relativistic contributions first
appearing in the third order are generally multiplied by a
$(\ell/\ell_H)^2$ factor which came from $\varphi$ terms in Eqs.\
(\ref{delta-eq-3rd})-(\ref{X-eq-3rd}), see $\varphi$ in Eq.\
(\ref{linear-sol-1}); $(\ell/\ell_H)^2$ is small in the small scale
but becomes of order unity in near horizon scale. The Newtonian part
of $\delta_3$ is proportional to $a^3$, and the relativistic part is
proportional to $a^2$. The Newtonian part of $\theta_3$ is
proportional to $a t$, and the relativistic part is proportional to
$a t^{1/3}$.

%
%
\section{Power spectra}

The density power spectrum is \bea
   & & | \delta ({\bf k}, t) |^2
       = | \delta_1 ({\bf k}, t) |^2
       + 2 {\cal R}
       \left[ \delta_1^* ({\bf k}, t) \delta_2 ({\bf k}, t) \right]
       + | \delta_2 ({\bf k}, t) |^2
       + 2 {\cal R}
       \left[ \delta_1^* ({\bf k}, t) \delta_3 ({\bf k}, t) \right]
       + \dots ,
\eea where ${\cal R}$ indicates the real part. The velocity power
spectrum $| \theta ({\bf k}, t) |^2$ can be similarly expanded. {\it
Assuming} the random phase, the second term in the right-hand-side
vanishes. The second-order power spectra of density and velocity
follow from Eqs.\ (\ref{linear-sol-1})-(\ref{third-order-sol-2}).
These are \bea
   & & | \delta ({\bf k}, t) |^2
       = | \delta_1 ({\bf k}, t) |^2
       + {1 \over (2 \pi)^3} \int d^3 k^\prime
       \Bigg\{
       {2 \over 14^2}
       | \delta_1 ({\bf k}^\prime, t) |^2
       | \delta_1 ({\bf k} - {\bf k}^\prime, t) |^2
       J^2 ( {\bf k}, {\bf k}^\prime,
       {\bf k} - {\bf k}^\prime )
   \nonumber \\
   & & \qquad
       +
       | \delta_1 ({\bf k}, t) |^2
       \bigg[
       | \delta_1 ({\bf k}^\prime, t) |^2
       \bigg(
       {2 \over 63}
       F ( {\bf k}, {\bf k}^\prime, {\bf k} - {\bf k}^\prime )
       L ( {\bf k} - {\bf k}^\prime, {\bf k}, - {\bf k}^\prime )
   \nonumber \\
   & & \qquad
       + {1 \over 18}
       H ( {\bf k}, {\bf k}^\prime )
       J ( {\bf k} - {\bf k}^\prime, {\bf k}, - {\bf k}^\prime )
       + {1 \over 18}
       H ( {\bf k}, {\bf k} - {\bf k}^\prime )
       L ( {\bf k} - {\bf k}^\prime, {\bf k}, - {\bf k}^\prime )
       \bigg)
       + C^+
       \bigg]
       \Bigg\}
   \nonumber \\
   & & \qquad
       + \,
       {10 \over 21}
       \left( {\ell \over \ell_H} \right)^2
       | \delta_1 ({\bf k}, t) |^2
       {1 \over (2 \pi)^3} \int d^3 k^\prime
       \Bigg\{
       | \delta_1 ({\bf k}^\prime, t) |^2
       \left(
       {13 \over 3}
       + 4 {k^2 \over k^{\prime 2}}
       + 7 { {\bf k} \cdot {\bf k}^\prime \over k^{\prime 2} }
       - 12 { ( {\bf k} \cdot {\bf k}^\prime )^2
       \over k^{\prime 4} }
       + 14 {k^2 \over k^{\prime 2}}
       { {\bf k} \cdot {\bf k}^\prime \over k^{\prime 2} }
       \right)
   \nonumber \\
   & & \qquad
       + \bigg[
       | \delta_1 ({\bf k}^\prime, t) |^2
       M ( {\bf k} - {\bf k}^\prime, {\bf k}, - {\bf k}^\prime )
       {k^2 \over |{\bf k} - {\bf k}^\prime|^2}
       \left(
       1 - 3 { {\bf k} \cdot {\bf k}^\prime
       \over k^{\prime 2} }
       + {15 \over 2}
       { {\bf k}^\prime \cdot ({\bf k} - {\bf k}^\prime)
       \over |{\bf k} - {\bf k}^\prime|^2 }
       + 3 {k^2 \over k^{\prime 2}}
       { {\bf k}^\prime \cdot ({\bf k} - {\bf k}^\prime)
       \over |{\bf k} - {\bf k}^\prime|^2 }
       \right)
       + C^+
       \bigg]
       \Bigg\},
   \label{P-density} \\
   & & | \theta ({\bf k}, t) |^2
       = {4 \over 9} (a^2/t^2) | \delta_1 ({\bf k}, t) |^2
       + (a^2/t^2) {1 \over (2 \pi)^3} \int d^3 k^\prime
       \Bigg\{
       {2 \over 21^2}
       | \delta_1 ({\bf k}^\prime, t) |^2
       | \delta_1 ({\bf k} - {\bf k}^\prime, t) |^2
       L^2 ( {\bf k}, {\bf k}^\prime,
       {\bf k} - {\bf k}^\prime )
   \nonumber \\
   & & \qquad
       +
       {2 \over 189} | \delta_1 ({\bf k}, t) |^2
       \bigg[
       | \delta_1 ({\bf k}^\prime, t) |^2
       \bigg(
       4 F ( {\bf k}, {\bf k}^\prime, {\bf k} - {\bf k}^\prime )
       L ( {\bf k} - {\bf k}^\prime, {\bf k}, - {\bf k}^\prime )
   \nonumber \\
   & & \qquad
       +
       H ( {\bf k}, {\bf k}^\prime )
       J ( {\bf k} - {\bf k}^\prime, {\bf k}, - {\bf k}^\prime )
       +
       H ( {\bf k}, {\bf k} - {\bf k}^\prime )
       L ( {\bf k} - {\bf k}^\prime, {\bf k}, - {\bf k}^\prime )
       \bigg)
       + C^+
       \bigg]
       \Bigg\}
   \nonumber \\
   & & \qquad
       + \,
       {20 \over 63}
       \left( {\ell \over \ell_H} \right)^2
       (a^2/t^2)
       | \delta_1 ({\bf k}, t) |^2
       {1 \over (2 \pi)^3} \int d^3 k^\prime
       \Bigg\{
       {2 \over 3}
       | \delta_1 ({\bf k}^\prime, t) |^2
       \left(
       {5 \over 3}
       + 8 {k^2 \over k^{\prime 2}}
       + 7 { {\bf k} \cdot {\bf k}^\prime \over k^{\prime 2} }
       - 24 { ( {\bf k} \cdot {\bf k}^\prime )^2
       \over k^{\prime 4} }
       + 14 {k^2 \over k^{\prime 2}}
       { {\bf k} \cdot {\bf k}^\prime \over k^{\prime 2} }
       \right)
   \nonumber \\
   & & \qquad
       + \bigg[
       | \delta_1 ({\bf k}^\prime, t) |^2
       M ( {\bf k} - {\bf k}^\prime, {\bf k}, - {\bf k}^\prime )
       {k^2 \over |{\bf k} - {\bf k}^\prime|^2}
       \left(
       {4 \over 3}
       - 4 { {\bf k} \cdot {\bf k}^\prime
       \over k^{\prime 2} }
       + 3
       { {\bf k}^\prime \cdot ({\bf k} - {\bf k}^\prime)
       \over |{\bf k} - {\bf k}^\prime|^2 }
       + 4 {k^2 \over k^{\prime 2}}
       { {\bf k}^\prime \cdot ({\bf k} - {\bf k}^\prime)
       \over |{\bf k} - {\bf k}^\prime|^2 }
       \right)
       + C^+
       \bigg]
       \Bigg\}.
   \label{P-velocity}
\eea Notice that, in the second-order power spectrum, compared with
the Newtonian contributions the pure general relativistic effects
are simply multiplied by a $(\ell/\ell_H)^2$ factor. Newtonian part
of the density spectrum is proportional to $a^4$ and the
relativistic part is proportional to $a^3$. In the case of the
velocity power spectrum, Newtonian part is proportional to $a^3$ and
the relativistic part is proportional to $a^2$.  Newtonian parts of
the power spectra are known in the literature
\cite{Vishniac-1983,quasilinear}. The pure relativistic
contributions to the second-order power spectra are our new
contribution in this work.

(i) For the power transfer from the small-scale, thus ${k}^\prime
\rightarrow \infty$, we have (in the following we {\it assume}
isotropic power spectrum, thus $| \delta ({\bf k}, t) | = | \delta
(k, t) |$, etc) \bea
   & & | \delta ({k}, t) |^2
       \simeq | \delta_1 ({k}, t) |^2
       + {1 \over 7^2 \pi^2} k^4
       \int {1 \over k^{\prime 2}} d k^\prime
       | \delta_1 ({k}^\prime, t) |^4
       - {1 \over 21 \pi^2}
       k^2 | \delta_1 ({k}, t) |^2
       \int d k^\prime
       | \delta_1 ({k}^\prime, t) |^2
   \nonumber \\
   & & \qquad
       - \left( {\ell \over \ell_H} \right)^2
       {130 \over 63 \pi^2}
       | \delta_1 ({k}, t) |^2
       \int k^{\prime 2} d k^\prime
       | \delta_1 ({k}^\prime, t) |^2.
\eea Using $\delta_k^2 \sim | \delta ({k}, t) |^2 k^3$, with
$\delta_k$ a density contrast at a given wavenumber $k$, and
introducing a small-scale cut-off frequency $k_c$ we have \bea
   & & | \delta ({k}, t) |^2
       \sim
       | \delta_1 ({k}, t) |^2
       + c_1 (k/k_c)^4 k_c^{-3} \delta_{k_c}^4
       + c_2 | \delta_1 ({k}, t) |^2 (k/k_c)^2 \delta_{k_c}^2
       + c_3 | \delta_1 ({k}, t) |^2
       (\ell / \ell_H)^2 \delta_{k_c}^2,
\eea where $c_i$ are constants of order unity. The $c_1$ term is the
well known $k^4$ long wavelength tail generated by the power
transfer from the small-scale nonlinearity; this was first shown by
Zel'dovich \cite{Zeldovich-1965,Vishniac-1983,textbook}. The
Newtonian and relativistic contributions in $2 {\cal R} (\delta_1^*
\delta_3)$ are smaller than the linear term by factors $(k / k_c)^2
\delta_{k_c}^2$ and $(\ell / \ell_H)^2 \delta_{k_c}^2$,
respectively; we have $k/k_c=\ell_c/\ell \ll 1$ with $\ell_c \equiv
a/k_c$.

(ii) For the power transfer from the large-scale, thus ${k}^\prime
\rightarrow 0$, we have \bea
   & & | \delta ({k}, t) |^2
       \simeq | \delta_1 ({k}, t) |^2
       + \left( {\ell \over \ell_H} \right)^2
       {5 \over 21 \pi^2}
       k^2 | \delta_1 ({k}, t) |^2
       \int d k^\prime
       | \delta_1 ({k}^\prime, t) |^2.
\eea To the lowest order in ${k}^\prime \rightarrow 0$, the
Newtonian contributions, which have order $ k^2 | \delta_1 ({k}, t)
|^2 \int d k^\prime | \delta_1 ({k}^\prime, t) |^2$, exactly cancel
out to the second order; this was shown by Vishniac
\cite{Vishniac-1983,quasilinear}. The next order terms in that limit
have order $| \delta_1 ({k}, t) |^2 \int k^{\prime 2} d k^\prime |
\delta_1 ({k}^\prime, t) |^2$. Thus, using a large-scale cut-off
frequency $k_c$ we have \bea
   & & | \delta ({k}, t) |^2
       \sim
       | \delta_1 ({k}, t) |^2
       + c_4 | \delta_1 ({k}, t) |^2 \delta_{k_c}^2
       + c_5 | \delta_1 ({k}, t) |^2
       (\ell / \ell_H)^2 (k/k_c)^2 \delta_{k_c}^2.
\eea Thus, the second-order terms, both Newtonian and relativistic,
are smaller than the linear ones by factors $\delta_{k_c}^2$ and
$(\ell / \ell_H)^2 (k/k_c)^2 \delta_{k_c}^2$, respectively. Compared
with the Newtonian second-order term the relativistic one has $\sim
({\ell / \ell_H})^2 (k/k_c)^2 \sim (\ell_c / \ell_H)^2$ which is
smaller than unity but has no scale dependence.

%
%
\section{Discussion}

Our results show that even to the second order the power spectra
receive corrections from pure relativistic third-order
perturbations. Compared with the Newtonian terms, the relativistic
contributions are multiplied by a factor $(\ell / \ell_H)^2$, thus
generally suppressed in the small-scale limit, but comparable in
near horizon scale. In near horizon scale, however, as the
perturbations are supposed to be in near linear stage, the
second-order contributions are negligible compared with the
linear-order terms. In the case of the power transfer from the
large-scale nonlinearity it was previously known that the leading
order Newtonian nonlinear contribution in the density power spectrum
cancels exactly. Even in that situation our investigation shows that
the relativistic effect is subdominant compared with the remaining
Newtonian nonlinear effect. Our analysis of the asymptotic cases
shows that the $k^4$ long wavelength tail of density power spectrum
generated by the small-scale nonlinearity in the Newtonian theory is
the only important effect of the nonlinear power transfer even in
Einstein's gravity.

Resolution of the issue of whether the relativistic contributions
are always smaller than the Newtonian nonlinear effects requires
quantitative estimation of the general power spectra presented in
Eqs.\ (\ref{P-density}) and (\ref{P-velocity}). This may depend on
the specific form of linear density power spectrum, and may require
numerical integration of Eqs.\ (\ref{P-density}) and
(\ref{P-velocity}). As the higher perturbational order contribution
to the power spectrum is in general suppressed by power of
$\delta_k^2$ term \cite{quasilinear}, it is likely that such
Newtonian contributions are smaller than the relativistic
contribution to the second order; quantitative estimation may depend
on the linear spectrum and the scale. Although the Newtonian
perturbation theory has a recursion formulae to all orders
\cite{quasilinear}, the relativistic situation should be handled at
each order separately. Estimation of non-Gaussian contribution of
the pure relativistic corrections is also an important and
interesting issue to be addressed: our Eqs.\
(\ref{delta-k-eq})-(\ref{X-k-eq}) or Eqs.\
(\ref{linear-sol-1})-(\ref{third-order-sol-2}) provide the starting
point for evaluating the bispectrum or the higher order correlation
functions. In the presence of the cosmological constant we have to
go back to Eqs.\ (\ref{delta-k-eq})-(\ref{X-k-eq}) which are valid
in the presence of the cosmological constant; for a general
expression of $\varphi$ in terms of $\delta$ and $\theta$ to the
linear order, see Eq.\ (\ref{varphi}). Our perturbation equations
are fully relativistic while assuming perturbations to be weakly
nonlinear. In the small scale where structures are in fully
nonlinear stage while the relativistic effects are small, the
cosmological post-Newtonian approximation provides a complementary
theoretical framework to handle the structure formation \cite{PN}.
Density power spectrum based on the cosmological post-Newtonian
equations has not been studied in the literature. Investigation of
these issues are left for future studies.

%
%
\subsection*{Acknowledgments}

JH was supported by the Korea Research Foundation Grant funded by
the Korean Government (MOEHRD, Basic Research Promotion Fund)
(KRF-2007-313-C00322). HN was supported by grants No.
R04-2003-10004-0 from the Basic Research Program of the Korea
Science and Engineering Foundation.

%
%

\end{document}